\documentclass[preprint]{aastex}

\newcommand{\be}{\begin{equation}}
\newcommand{\ee}{\end{equation}} 
\newcommand{\sigwig}{{ \langle {\widetilde\sigma} \rangle }}  
\newcommand{\sigbar}{{\langle \sigma \rangle}}  
\newcommand{\sigcap}{{ \langle \sigma_{\rm cap} \rangle }}  
\newcommand{\zfactor}{{{\cal F}_0}} 
\newcommand{\nstar}{{N_\star}} 
\newcommand{\nrock}{{n_{\rm rock}}} 
\newcommand{\ncap}{{N_{\rm cap}}} 
\newcommand{\nbio}{{N_{\rm bio}}} 
\newcommand{\numrelax}{{{\cal F}_R}} 
\newcommand{\tauwig}{{ {\widetilde \tau} }} 
\newcommand{\twig}{{ {\tilde t} }} 
\newcommand{\fimp}{{f_{\rm imp}}} 
\newcommand{\cfraction}{ f_{\rm cap}} 
\newcommand{\mearth}{{ M_\oplus }} 
\newcommand{\vkms}{ v_{\rm kms}} 
\newcommand{\veject}{ v_{\rm eject}} 
\newcommand{\vcluster}{ v_{\rm cls}}  
\def\lta{\,\raise 0.3 ex\hbox{$ < $}\kern -0.75 em
 \lower 0.7 ex\hbox{$\sim$}\,}
\def\gta{\,\raise 0.3 ex\hbox{$ > $}\kern -0.75 em
 \lower 0.7 ex\hbox{$\sim$}\,} 
\begin{document}


\title{\bf Lithopanspermia in Star Forming Clusters}  

\author{Fred C. Adams$^{1,2}$ and David N. Spergel$^3$} 
 
\affil{$^1$Michigan Center for Theoretical Physics, University of Michigan \\
Physics Department, Ann Arbor, MI 48109}  

\affil{$^2$Astronomy Department, University of Michigan, Ann Arbor, MI 48109}

\affil{$^3$Department of Astrophysical Sciences, Princeton University \\ 
Princeton, NJ 08544} 



\begin{abstract} 

This paper considers the lithopanspermia hypothesis in star forming
groups and clusters, where the chances of biological material
spreading from one solar system to another is greatly enhanced
(relative to action in the field) due to the close proximity of the
systems and lower relative velocities.  These effects more than
compensate for the reduced time spent in such crowded environments.
This paper uses $\sim300,000$ Monte Carlo scattering calculations to
determine the cross sections $\sigcap$ for rocks to be captured by
binaries and provides fitting formulae for other applications. We
assess the odds of transfer as a function of the ejection speed
$\veject$ and number $\nstar$ of members in the birth aggregate. The
odds of any given ejected meteroid being recaptured by another solar
system are relatively low, about 1 in $10^3 - 10^6$ over the expected
range of ejection speeds and cluster sizes. Because the number of
ejected rocks (with mass $m > 10$ kg) per system can be large, $N_R
\sim 10^{16}$, virtually all solar systems are likely to share rocky
ejecta with all of the other solar systems in their birth cluster. The
number of ejected rocks that carry living microorganisms is much
smaller and less certain, but we estimate that $N_B \sim 10^7$ rocks
can be ejected from a biologically active solar system. For typical
birth environments, the capture of life bearing rocks is expected to
occur $\nbio \approx$ 10 -- 16,000 times per cluster (under favorable
conditions), depending on the ejection speeds. Only a small fraction
($\fimp \sim 10^{-4}$) of the captured rocks impact the surfaces of
terrestrial planets, so that $N_{\rm lps} \approx 10^{-3} - 1.6$
lithopanspermia events are expected (per cluster). Finally, we discuss
the question of internal vs external seeding of clusters and the
possibility of Earth seeding young clusters over its biologically
active lifetime.
 
\end{abstract}



\section{INTRODUCTION} 

The question of whether life can be spread from one solar system to
another is of long standing interest to astrobiology. Previous
estimates (e.g., Melosh, 2003) suggest that the odds of both meteroid
and biological transfer are exceedingly low. However, such estimates
have been determined for the conditions in the local galactic
neighborhood, i.e., in the field. The odds of transfer increase in
more crowded environments.  Although the vast majority of stars
currently reside in the field, most stars form within small groups or
clusters, rather than in isolation (e.g., Lada and Lada, 2003; Porras
{\it et al.}, 2003; see also Adams and Myers, 2001). Since the time
scale for planet formation and the time that young stars are expected
to live in their birth clusters are roughly comparable, about 10 -- 30
Myr, debris from planet formation has a good chance of being
transferred from one solar system to another.  A related question is
whether or not biologically active material can be transferred from
one system to another. Because of the destructive effects of
ultraviolet (UV) radiation, and other hardships of deep space, current
thinking suggests that biological material must be encased in rock in
order to survive the transfer process. The required rock mass is often
taken to be $m > 10$ kg (Horneck 1993; Nicholson {\it et al.}, 2000;
Benardini {\it et al.}, 2003; Melosh, 2003) although better transfer
odds can be obtained if this mass is smaller (Napier, 2004). As a
result of the required UV shielding, this mechanism is generally
called lithopanspermia.  Suppose that life can be seeded into one
solar system in a young cluster, either by direct biogenesis or
through a chance encounter with bio-invested material from outside the
birth aggregate. Subsequent dynamical interactions among the
constituent solar systems can then allow life to spread throughout the
birth cluster. The goal of this work is to estimate the likelihood of
such transfer.

The transfer of rocks between planets within a solar system is a well
established phenomena. Researchers on Earth regularly find and study
Martian meteorites (McSween, 1985), and the dynamics of such transfer
has been well studied (e.g., , Gladman, 1997; Dones {\it et al.}, 1999; 
Mileikowsky {\it et al.}, 2000).  The transfer of rocky material between
solar systems is more difficult, but should still take place. The
exchange of life bearing meteroidites between solar systems is unlikely
to occur between field stars (Melosh, 2003) due to the high relative
velocities and low stellar density.  As we show here, however, in
young stellar groups and clusters, where most stars form, the stellar
densities are higher and the relative velocities are lower. These
properties increase the odds of transfer.  Furthermore, previous
estimates of capture cross sections have focused on single stars; most
stars live in binary systems and the capture cross sections for
binaries are greatly enhanced.  

In this paper, we present a comprehensive evaluation of the
lithopanspermia mechanism in star forming groups and clusters. We
first perform a series of numerical calculations to estimate the
distributions of ejection speeds for rocks exiled from their solar
systems and discuss the distribution of rock sizes and masses. We
next consider the dynamics of young groups and clusters and find the
optical depth for ejected rocks to be recaptured by other solar
systems. In order to make such estimates, we calculate the capture
cross sections using a Monte Carlo technique and a large ensemble
($\sim300,000$) of numerical experiments to sample the parameter
space. Putting all of these components together, we estimate the
expected number of rocks to be recaptured, the expected number of life
bearing rocks to be transferred, and finally the expected number of
successful lithopanspermia events (per cluster).  Note that clusters
are not guaranteed to have any of their member solar systems develop
life during the 10 -- 100 Myr that they remain bound. As a result,
this paper calculates the odds of lithopanspermia events only for
those clusters that produce at least one living system.

\section{LITHOPANSPERMIA IN GROUPS AND CLUSTERS} 

\subsection{Ejection of rocky bodies from a solar system} 

One can think of each solar system in the cluster as a source of rocky
debris. In other words, each solar system produce a mass outflow rate
of rocky material. Our present solar system contains about 50 Earth
masses ($\mearth$) of rocky bodies, with most of the mass residing in
the cores of the Jovian planets. The early solar nebula contained 50
-- 100 $\mearth$ of heavy elements.  Here, we parameterize the heavy
element content $M_Z$ of nascent solar systems as 
\be 
M_Z = \zfactor \mearth \, , 
\ee
where we expect $\zfactor$ = 50 -- 200 for typical systems. 

Since planets form within solar systems on time scales of roughly
$t_p$ = 10 Myr (Lissauer, 1993), and since some fraction $f_e$ of the
rocky material will be ejected during the course of planet formation 
and subsequent dynamical evolution of the system, the mean mass loss 
rate (in rocks) from a young solar system is given by 
\be 
\langle {dM \over dt} \rangle = {f_e \zfactor \mearth \over t_p} \, . 
\ee 
Numerical studies (Dones {\it et al.}, 1999; see also Melosh, 2003)
suggest that about one third of the material not locked up in planets
will be ejected from a given solar system containing giant planets
(such as Jupiter and Saturn).  Since planet formation is unlikely to
be 100 percent efficient, we might expect one third of the initial
material to be left over, and one third of that to be ejected (with a
good fraction of the remainder accreted by the central star), so that
$f_e \approx 1/10$. Since the times scales for planet formation are
roughly the same as the lifetimes for small stellar clusters (Binney
and Tremaine, 1987), this process has time to grind towards completion
while the cluster remains intact, and hence each solar system
contributes a mass $M_R$ = $f_e \zfactor \mearth$ of rocky material to
its birth aggregate. A conservative benchmark value for this mass
scale is thus $M_R \approx \mearth$, with a corresponding mass loss
rate of $\langle dM /dt \rangle \approx 10^{-7}$ $\mearth$ yr$^{-1}$
$\approx$ $2 \times 10^{13}$ g s$^{-1}$. This mass loss rate may seem
large.  For comparison, the mass equivalent loss due to solar
radiation escaping from our solar system is ${\dot m} = L_\odot/c^2
\approx 4 \times 10^{12}$ g s$^{-1}$, only about five times smaller.

The speeds of rocks ejected from a solar system depend on their
initial location (the depth of the stellar gravitational potential
well) and the mass of the scattering body (the depth of its
gravitational potential well). When biologically active rocks are
removed from the surface of their parent planet through impacts, they
can either be directly ejected from the solar system or be left in
orbit (about the central star) where they are subsequently ejected by
other solar system bodies.  As representative examples of this latter
process, we have performed three ensembles of scattering calculations
to sample the possible ejection speeds. In these 3-body experiments, a
small rocky body and a companion (either a giant planet or star) are
placed in orbits about a primary star with mass $M_\ast = 1.0
M_\odot$. The companion is taken to have a moderately eccentric orbit,
whereas the rocks are given initial semi-major axes and eccentricities
so that the orbits bring the bodies near each other (sometimes
orbit-crossing).  Specifically, the rocks have eccentricities randomly
drawn from the interval [0,0.5] and semi-major axes $a = \xi a_C$,
where $a_C$ is the semi-major axis of the companion and $\xi$ is a
log-random variable selected from the range $\log_{10}\xi \in
[-1,1]$. The simulations are co-planar and are integrated using a
Bulirsch-Stoer scheme.

The resulting distributions of ejection speeds are shown in Figure 1
for three representative cases: a Jupiter-like planet (with the mass
of Jupiter and semi-major axis $a_C$ = 5 AU), a Neptune-like planet
(with the mass of Neptune and $a_C$ = 30 AU), and a 0.1 $M_\odot$ binary
companion with $a_C$ = 42 AU (near the peak of the binary period
distribution). The distribution of ejection speeds is similar for the
two planetary cases, with median values $\veject$ = 5.4 km/s (5.8
km/s) and mean values $\langle \veject \rangle$ = $6.2 \pm 2.7$ km/s
($6.1 \pm 1.3$ km/s) for giant planet analogs of Jupiter (Neptune).
These results are in good agreement with previous work that found
ejection speeds of $\veject \approx 5 \pm 3$ km/s for rocky bodies
scattering out of a solar system due to perturbations from Jupiter
(Melosh, 2003). The distribution of ejection speeds for the stellar
companion has a somewhat smaller median ($\veject$ = 4.8 km/s) and
mean value ($\langle \veject \rangle = 5.0 \pm 2.5$ km/s); more
significantly, the distribution is wider and has substantial support
at smaller ejection speeds. Additional simulations (not shown here)
indicate that rocky ejecta from solar systems with smaller primaries
will generally have lower ejection speeds. We note that solar systems
can have a wide variety of architectures (e.g., Levison, Lissauer, and
Duncan, 1998; Levison and Agnor, 2003), including planetary systems
encircled by binary companions (e.g., David {\it et al.}, 2003), so
that an even wider distribution of ejection speeds remains possible.

\subsection{Distribution of rock sizes and masses} 

For a given mass in rocky material, we need to specify its mass
distribution. The distribution of interplanetary bodies has been
discussed previously and the differential mass distribution generally
takes a power-law form 
\be 
{dN \over dm} = B m^{-\alpha} \, ,
\label{eq:dist} 
\ee
where $B$ is the normalization constant. The slope $\alpha$ has a
canonical value of about 1.83 for systems dominated by collisions
(see, e.g., Hughes and Daniels, 1982; Napier 2001) and about 1.67 for
rocks hitting the atmosphere of Earth (Schroeder, 1991). We assume
here that the distribution of rocky bodies in a forming solar system
has this general form, although the normalization (which sets the
total mass in rocky material) can vary from system to system. The
distribution can be normalized by requiring a fixed total mass $M_R$
in rocky bodies, i.e., 
\be 
M_R = B \int dm \, m^{-(\alpha-1)} \, . 
\ee
In order to keep the integral from diverging, we introduce 
an upper mass cutoff $m_2$ and thereby obtain 
\be 
B = (2 - \alpha) \, M_R \, m_2^{-(2-\alpha)} \, . 
\ee 
Similarly, in order to keep the number of rocky bodies from diverging,
we must impose a lower mass cutoff $m_1$. For distributions of this
type, namely with indices in the range $1 < \alpha < 2$, essentially
all of the mass resides in the upper end of the range, whereas all of
the rocky bodies, by number, reside in the lower end of the range.
In this setting, we are only interested in bodies larger than the 
minimum size/mass required to shield biological material. This lower 
mass limit $m_1$ is often taken to be 10 kg (Melosh, 2003), although 
alternate values have been suggested (Napier, 2004). For given values 
of the upper and lower mass scales, the total number $N_R$ of bodies 
with mass $m > m_1$ thus becomes 
\be 
N_R = {2- \alpha \over \alpha - 1} 
{M_R \over m_1^{(\alpha-1)} m_2^{(2-\alpha)} } \, .  
\ee 
For one set of typical values --- $\alpha$ = 5/3, $m_1$ = 10 kg, 
$m_2$ = 0.1 $\mearth$, and $M_R$ = $\mearth$ (Melosh, 2003) ---  
the total number of rocky bodies becomes $N_R \approx 10^{16}$. 
For an alternate set of values --- $\alpha$ = 11/6 and $m_1$ = 
$10^4$ kg (Napier 2004) --- we obtain $N_R \approx 8 \times 10^{16}$. 

\subsection{Life bearing rocks} 

Only some fraction of the material ejected from the solar system will
be biologically active (seeded with spores or other biological
material). We denote this fraction as $f_B$, so that the total mass
(per system) of biologically active material is $f_B N_R$. Although
the fraction $f_B$ is not well determined, it is significant that many
bacteria -- those known as extremophiles -- are well suited to
survival in harsh conditions. In particular, the bacterium {\it
Deinococcus radiodurans} seems almost designed for space travel -- it
can withstand extreme doses of radiation, cold, oxidation damage, and
can survive for long periods without water (e.g., Minton, 1994;
Battista, 1997; White {\it et al.}, 1999).

Previous papers have estimated the minimum mass $m_B$ required for
biologically active material to survive in deep space and found that
$m_B \approx 10$ kg (Horneck, 1993; Nicholson {\it et al.}, 2000);
this value motivates our choice of lower mass cutoff $m_1 = m_B$. As a
result, the maximum number of biologically active units that could be
provided by a single solar system would be $N_R$ if all the rocks were
biologically active. As discussed above, $N_R \sim 10^{16}$, although
this number is uncertain and varies from system to system. Nonetheless,
this value provides a good starting point.

Unfortunately, the effectiveness of panspermia depends sensitively on
the lower mass limit $m_B$ for life to survive in rocks. If the mass
scale is much lower (e.g., Napier, 2004), then the steep distribution
function (equation [\ref{eq:dist}]) implies an enormous increase in
the number of available missiles. On the other hand, life bearing
rocks that are captured by other solar systems and land on terrestrial
planets are subject to another peril: The micro-organisms must survive
the landing. Small rocks burn up in the atmosphere and/or reach
temperatures too high for life to survive. Larger rocks will make it
through the atmosphere and can have interior temperatures low enough
for life to survive, but such large rocks tend to have violent impacts
with the planetary surface. These impacts can also heat up the rocks
and cause the destruction of biological material contained within.
Some data on this issue exists: The observed mass distributions of
meteorites on the Antarctic ice (Huss, 1990) do not flatten out for
masses greater than about 100 grams, which implies that the meteors
larger than 100 grams have a good chance of surviving their fall
through the atmosphere. The 10 kg rocks considered here should thus be
safe (see also Wells, Armstrong, and Gonzales, 2003).  Microorganisms
in larger rocks have an even greater chance of survival. Meter-sized
bodies reach modest terminal speeds (for an atmosphere of terrestrial
density) and experimental data indicate that bacteria can survive the
landing (e.g., Burchell et al., 2001, 2004; Mastrapa et al., 2001).

We also need an estimate of the fraction $f_B$ of ejected rocks that
carry biological material. Suppose, for example, that life is seeded
on a large terrestrial planet in the system. It is reasonable to
suppose that life would quickly spread over the surface and into the
planet down to depth $\ell$ of a few kilometers.  As an optimistic
benchmark scenario, we can assume that the mass $3 (\ell/R_E) \mearth
\sim 10^{-3} \mearth$ is biologically active and is blasted away from
the surface by the intense early bombardment phase of that solar system. 
If this mass is broken up into $m_B$ = 10 kg pieces, then the number
of life bearing rocks would be $N_B \sim 10^{20}$. However, it would
be more likely for the mass to be broken up into a range of sizes,
e.g., distributed according to the considerations of the previous
section. In this case, the number of biologically active rocks would
be $N_B \approx 1.5 (\ell/R_E) \mearth / (m_1^2 m_2)^{1/3} \sim
10^{13}$.

The number of biologically active rocks has been estimated previously
(e.g., Melosh, 2003; Wallis and Wickramasinghe, 2004).  As another
benchmark, Melosh (2003) estimates that about 15 rocks per year (above
the minimum mass $m_B$) should be ejected from the surfaces of
terrestrial planets due to impacts (see also Melosh and Tonks, 1994;
for a more detailed discussion of boulder ejection, see Wallis and
Wickramasinghe, 2004).  Over the time scale of 10 Myr considered here,
a terrestrial planet with life would contribute $N_B \sim 10^7$ life
bearing rocks. Taken together, these considerations suggest that the
number of life bearing rocks lies in the range $N_B = 10^7 - 10^{14}$,
with the lower end of the range being strongly favored. We take $N_B =
10^7$ as our standard value for the remainder of this paper.  Notice,
however, that a smaller minimum mass $m_B$ (for biological protection)
implies a larger number of rocks $N_R$ and a larger number of life
bearing rocks $N_B$. Since both values scale proportional to
$m_B^{-2/3}$, if $m_B$ were as small as 142 grams (the mass of a
baseball), then the number of life bearing rocks would be larger by a
factor of $\sim17$.  On the other hand, during the late heavy
bombardment of Earth, large impacts may have led to a sterilization of
the planet, at least on the surface. As a result, the number of
biologically active rocks could be much smaller than the numbers
quoted above.

\subsection{Dynamical scattering interactions in groups and clusters} 

Stars -- and hence solar systems -- form in groups and clusters with a
range of sizes, but a large fraction are born in stellar aggregates
within the size range $\nstar$ = 100 -- 1000.  A cluster with $\nstar$
members has a typical radius of $R = R(\nstar)$ = 1 pc
$(\nstar/100)^{1/2}$, where this formula follows from a fit to the
data presented in Lada and Lada (2003) and Carpenter (2000). The
average starting number density $n_{\star 0}$ of solar systems in the
birth cluster is given by 
\be
n_{\star 0} = {\nstar \over 4 (\pi/3) R^3} \approx 
{750 \, {\rm pc}^{-3} \over \pi \nstar^{1/2} } \, . 
\ee
The effective ``scattering optical depth'' $\tau$ for interactions
between a passing body and a member of the cluster is given by the
integral 
\be 
\tau = \int n_\star \sigbar v dt \, , 
\ee
where $\sigbar$ is the cross section for interaction -- in this case,
the cross section for a solar system to capture a passing piece of rock. 

The scattering optical depth depends on the speeds at which rocks are
ejected from their solar systems. For ejection speeds less than (or
comparable to) the velocity dispersion of the cluster, the rocks are
dynamically bound and will orbit within the gravitational potential
well of the cluster for many crossing times. In the opposite limit,
high speed rocks only experience a single crossing time before passing
out of the cluster.  As shown in \S 2.1, rocky bodies scattered by
Jupiter and Neptune are expected to have ejection speeds $\veject
\approx 3 - 9$ km/s, somewhat larger than the typical velocity
dispersion $\vcluster \approx 1$ km/s for a small cluster. As a
result, many of the rocks will reside in the high speed regime with $v
\sim 5$ km/s. Because the solar systems themselves are moving with
relative speeds $v \approx \vcluster \sim 1$ km/s, and because the
scattering optical depth is a decreasing function of $v$, using the
velocity scale $\vcluster$ results in the largest possible scattering
optical depth $\tau$, i.e., an upper limit on the efficacy of
transfer.

For rocks that are bound to the cluster, we can ``evaluate'' the 
optical depth integral by writing it in the form  
\be 
\tau = \numrelax n_{\star 0} v_0 t_{R0} \sigbar \, , 
\label{eq:taudef} 
\ee
where the subscript `0' refers to the values at the beginning of the
cluster's life. The time scale $t_{R0}$ is the initial value of the
dynamical relaxation time. The total effective lifetime of the cluster
is then given by the time scale $\numrelax t_{R0}$. Dynamical studies
(see Binney and Tremaine, 1987) indicate that clusters have total
lifetimes of 50 -- 100 times the initial relaxation times. During this
time, however, the number density $n_\star$ and typical speeds $v$ of the
solar systems decrease substantially. We thus need to take lifetime
factor $\numrelax$ to be somewhat smaller, $\numrelax \approx$ 10. The
starting relaxation time is given by 
\be
t_{R0} = {R \over v_0} {\nstar \over 10 \log \nstar} \, , 
\ee 
and the effective optical depth of interaction becomes 
\be 
\tau \approx {3 \numrelax \nstar^2 \sigbar \over 40 \pi R^2 \log \nstar} 
\approx 24 {\nstar \over \log \nstar} {\sigbar \over (1{\rm pc})^2} \, . 
\label{eq:taulow} 
\ee 
This quantity will generally be less than unity, and thus represents
the probability that a given piece of rock will be captured by some
solar system in the cluster, during the time interval for which the
cluster remains intact. 

A related quantity is the optical depth of interaction $\tauwig$ for 
a given solar system to capture any piece of rock from an alien solar 
system. This second optical depth is given by the same integral form 
\be 
\tauwig = \int \nrock \sigbar v dt \, , 
\ee 
where $\nrock$ is the number density of rocky bodies that the 
solar system encounters. The remaining quantities are the same as 
before (equation [\ref{eq:taudef}]). In this case, we assume that 
each solar system ejects (on average) a given number $N_R$ of rocks, 
and that the velocity distribution of these rocky bodies follows that 
of the stars (which should be the case for low speed ejections since 
both populations are living in the same gravitational potential well). 
As a result, one expects that $\nrock \approx (N_R/\nstar) n_\star$
and hence $\nstar \tauwig = N_R \tau$, where this latter quantity is
the expected number of capture events for the entire group/cluster.

For rocks with higher initial velocities, the total path length 
sampled by a passing rock is of order one crossing length, i.e., 
$\int v dt \approx R$. The scattering optical depth is thus 
given by 
\be 
\tau = n_\star \sigbar R \, = {3 \nstar \sigbar \over 4 \pi R^2} 
\approx 24 {\sigbar \over (1 {\rm pc})^2} \, . 
\label{eq:tauhigh} 
\ee 
Because of the manner in which the cluster sizes $R$ scale with
stellar membership $\nstar$, the scattering optical depth $\tau$ is
nearly independent of the cluster richness. This optical depth in the
high speed limit is smaller than that of the low speed limit by a
factor of $\nstar/\log\nstar$ and also has a smaller cross section
(which is a sharply decreasing function of $\veject$ -- see the
following section).

\subsection{Interaction cross sections} 

The optical depths for interactions derived above can be applied 
to a wide variety of events provided that the cross section for 
the event is known. In this context, we are interested in two 
separate but related issues: The capture of passing rocky bodies 
by other solar systems (which are mostly binaries), as well as 
the possibility that the rocky body strikes the surface of a 
terrestrial planet. The cross section for this process can thus 
be written in the form 
\be 
\sigbar = \sigcap \fimp \, , 
\ee  
where $\sigcap$ is the capture cross section and $\fimp$ is 
the fraction of captured rocks that strike the surface of a 
terrestrial planet in the system. 

We have calculated the capture cross sections $\sigcap$ using a
scattering code developed previously (Adams and Laughlin, 2001; 
Laughlin and Adams, 2000) to study the dynamics of solar systems
interacting with binaries (most star systems are binary -- see Abt, 
1983). In this context, we perform a series of calculations to the
study the capture of rocky bodies by binary star systems.  Individual
encounters are treated as 3-body problems in which the equations of
motion are integrated using a Bulirsch-Stoer scheme. We separate out
the semi-major axis of the binary from the other variables (see below)
and write the capture cross section $\sigcap$ in terms of the integral 
\be 
\sigcap \equiv \int_{0}^{\infty} \cfraction 
(a) (4 \pi a^{2}) p(a) \, da \, , 
\ee 
where $a$ is the semi-major axis of the binary orbit and $p(a)$ is the
distribution of $a$ (determined from the observed distribution of
binary periods -- see Kroupa, 1995). This treatment includes only those
interactions within the predetermined area $4 \pi a^{2}$ (more distant
encounters are neglected because they have little effect). The function
$\cfraction (a)$ represents the fraction of encounters that result in
capture.

Dynamical encounters between a given rock and a field binary are
described by 10 input parameters (see Laughlin and Adams, 2000).
These variables include the binary semi-major axis $a$, the stellar
masses $m_{\ast 1}$ and $m_{\ast 2}$, the eccentricity $\epsilon_{\rm
b}$ and the initial phase angle $\ell_{\rm b}$ of the binary orbit,
the asymptotic velocity $v_\infty$ of the rock relative to the center
of mass, the angles $\theta$, $\psi$, and $\phi$ which describe the
impact direction and orientation, and finally the impact parameter $h$
of the encounter.

To compute the fraction of captures $\cfraction (a)$ and the
corresponding cross sections, we perform a large number of numerical
experiments using a Monte Carlo scheme to select the input
parameters. The binary eccentricities are sampled from the observed
distribution (Duquennoy and Mayor, 1991).  Masses of the two binary
components are drawn separately from an initial mass function (IMF)
consistent with the observed IMF (in particular, the form advocated by
Adams and Fatuzzo, 1996). The impact parameters $h$ are chosen randomly
within a circle of radius $2a$ centered on the binary center of mass.
The impact velocities at infinite separation $v_\infty$ are sampled
from a Maxwellian distribution with a given dispersion $\sigma_v$.
Here, we calculate the cross sections as a function of the dispersion
$\sigma_v$. For low ejection speeds, the rocks are bound to the
cluster and the relevant velocity dispersion is determined by the
gravity of the cluster ($\sigma_c \approx \vcluster \approx 1$ km/s). 
For higher ejection speeds, the rocks are not necessarily bound and
the relevant velocity dispersion is given by the dynamics of the
ejection process (see Figure 1).

Using the methodology described above, we have performed approximately
300,000 numerical experiments to sample the parameter space.  The
resulting cross sections are shown in Figure 2 as a function of the
velocity dispersion $\sigma_v$. Figure 2 also presents a fit to
the cross sections, where the fitting function has the form 
\be 
\sigcap = (51,900 \, {\rm AU}^2) \, \, \vkms^{-1.79} \, 
\exp[ -0.235 (\ln \vkms)^2] \, , 
\label{eq:fit} 
\ee 
where $\vkms$ is the velocity dispersion in units of km/s. 

The capture cross sections calculated here are much greater than those
used in the previous study of Melosh (2003) for two reasons: (1) Most
stars reside in binary systems and binarity increases the interaction
cross sections. The work of Melosh (2003) uses a Jupiter-mass
companion, although the cross sections increase with companion mass;
an extrapolation of those results leads to estimates compatible with
those calculated here (see also Laughlin and Adams, 2000; Adams and
Laughlin, 2001). (2) The cross sections are sensitive to the relative
velocity of the interacting systems. In a young cluster this velocity
scale is only $\sigma_v \sim 1$ km/s and the ejection speeds $\veject
\sim 5$ km/s; both values are much smaller than the velocity
dispersion of field stars where $\sigma_v \sim 20 - 40$ km/s.

Although binarity increases the interaction cross sections, binary
systems have a lower probability of supporting stable planetary orbits
in their habitable zones and the companion can inhibit terrestrial
planet formation. These effects are surprisingly modest: Over 50
percent of binary systems are wide enough to allow for Earth-like
planets to remain stable over the current 4.6 Gyr age of the solar
system (David {\it et al.}, 2003). The reason for this large fraction
of viable systems is that most binaries are wide, with the peak of the
binary period distribution at $P_b \approx 10^5$ days (Duquennoy and
Mayor, 1991). Such orbits are wide enough ($a \sim 42$ AU for a solar
mass primary) to allow stable orbits in the terrestrial region, as
well as a stable `Jupiter' with semi-major axis $a$ = 5 AU. For this
same reason, most binary companions do not inhibit the formation of
terrestrial planets (Quintana, 2004; Quintana {\it et al.}, 2002).

\subsection{Transfer probabilities} 

Given the above considerations, we can now evaluate the odds for rocks
to be ejected by one solar system and captured by another. The odds of
life bearing rocks being transferred can be determined similarly.
Note that the results depend rather sensitively on the ejection speeds
for rocks expelled from a solar system. In order to cover the range of
possibilities, we discuss both the low speed limit and the high speed
limit. In the low speed limit, defined by when the ejection speed
$\veject$ is less than the stellar velocity dispersion $\vcluster$ in
the cluster, the relative speed of interaction between rocks and solar
systems is determined by the stellar motions and hence $v = \vcluster
\sim 1$ km/s. In the high speed limit, the ejection speeds are larger
than relative speeds between solar systems, and $\veject$ determines
the relative speeds for capture interactions ($\veject \approx 5$ km/s
for many solar systems as illustrated by Figure 1). Because $v \approx$ 
1 km/s represents an upper limit on the scattering optical depth (see 
\S 2.4), we define a fiducial cross section $\sigbar_1 \equiv 52,000$ 
AU$^2$ $\approx \pi (129 {\rm AU})^2$, as found numerically for
$\sigma_v$ = 1 km/s. In order to allow for easy scaling of our
results, we also define a reduced cross section $\sigwig \equiv
\sigcap / \sigbar_1$. The velocity dependence of $\sigwig$ is given by
equation (\ref{eq:fit}) and by Figure 2.

In the low speed limit, the effective optical depth of interaction
(per rock) becomes  
\be 
\tau \approx (3 \times 10^{-5}) \, 
{\nstar \over \log \nstar} \, \sigwig \, \, , 
\label{eq:taulowev} 
\ee  
where we expect $\sigwig \approx 1$. The typical stellar population
for a clustered star formation region is about $\nstar=300$, so $\tau
\approx 0.0016$, and the number of rocks needed to get a capture event
is about $\tau^{-1} \approx 630$. 

In the high speed limit, the rocks only stay in the cluster for one
crossing time and the effective optical depth for interactions becomes 
\be 
\tau \approx (3 \times 10^{-5}) \sigwig \, , 
\label{eq:tauhighev} 
\ee
where $\sigwig \ll 1$. For a typical ejection speed $\veject$ = 5 km/s, 
for example, the reduced cross section $\sigwig \approx$ 0.0305 and
the scattering optical depth is $\tau \approx 10^{-6}$. In other
words, only about one out of a million rocks are recaptured.

The total number of capture events in the entire cluster is given by
$\ncap = N_R \tau$, where $N_R \sim 10^{16}$ (see \S 2.2).  For the
low speed limit $\ncap \approx 10^{13}$, and for the high speed case
with $\veject = 5$ km/s, $\ncap \approx 10^{10}$.  As a result, the
solar systems in a typical birth aggregate will experience billions to
trillions of capture events, where ``capture events'' are the capture
of rocky bodies from a single given solar system. It is possible --
and even likely -- that every solar system will contribute $N_R \sim
10^{16}$ rocky bodies to the cluster environment. As a result,
essentially every solar system in a cluster can share rocky material
with all of the other solar systems in its birth aggregate.

The number of life bearing rocks is far lower than the total and has a
large uncertainty. Since the origin of life is presumably a rare event
in contexts where panspermia is of interest, we expect that at most
one solar system would (initially) become biologically active and
capable of seeding the rest of its birth cluster. We thus consider
only one system as the source of bioactive rocks.  Keep in mind,
however, that not all clusters are guaranteed to develop life. As a
result, the number of successful lithopanspermia events calculated
here should be multiplied by the fraction $f_{cl}$ of clusters that
contain at least one living system (independent of panspermia).  For
our benchmark value $N_B = 10^7$, the number of captured life bearing
rocks is $\nbio \approx$ 16,000 over the entire cluster in the limit
of low ejection speeds.  On average, every solar system would capture
about 50 life bearing rocks from the parent system.  For higher
ejection speeds (here, $\veject = 5$ km/s), $\nbio \approx 10$ and
only 1 out of 30 solar systems in a typical birth cluster are expected
to capture biologically active rocks.  Nonetheless, some transfer of
life bearing rocks is likely to occur within young star clusters.

Although the capture of life bearing rocks is necessary to spread life
from one planet to another, it is not sufficient. A captured rock must
eventually find its way from its initial orbit (that resulting from
the capture process) to the surface of a suitable terrestrial planet. 
The probability $\fimp$ for a captured rock to strike a terrestrial
planet is generally very small. Melosh (2003) has performed a series
of simulations to estimate this quantity and finds that the
probability of impact over the entire age of our solar system (4.5
Gyr) is only about $\fimp \sim 10^{-4}$. The probability of impact on
a large rocky moon, in orbit about a giant planet, is somewhat smaller, 
$\fimp \sim 10^{-5}$. As a working benchmark value, we adopt $\fimp
\sim 10^{-4}$. 

In this setting, however, the biologically active rocks are often
captured while the solar systems are young and hence still in the
process of building planets. These systems are extremely active and
collisions are common. The life bearing rocks have a much better
chance (compared with the case of mature solar systems) of colliding
with other debris and infecting them with spores. Although a life
bearing rock will collide with many other rocky bodies, the efficiency
of transfer (from rock to rock) is not known. In the long run, a large
fraction of the total rocky content of a solar system will become
incorporated into surviving bodies -- giant planets, moons,
terrestrial planets, or asteroids -- so any infected rocks have a good
chance of seeding life on larger bodies.  This effect will act to make
the quantity $\fimp$ larger than that calculated previously for more
mature solar systems. Our adopted fiducial value ($\fimp \sim
10^{-4}$) should thus be considered as a lower limit to the transfer
efficiency.  In addition to rocky bodies, comets provide another
useful vehicle for the transfer of biologically active material. As
an added advantage, when comets pass through the atmosphere of an
Earth-like planet, they tend to disintegrate into dust and can thereby
deposit biological material in a viable state (e.g., Narlikar et
al., 2003). For further discussion regarding the survival of
microorganisms during infall, see Hoyle et al. (1999).

As described above, the expected number of captured rocks that are
potentially biologically active is $\nbio$ = 10 -- 16,000 per cluster, 
depending mostly on the ejection speeds. Over this range, the expected
number of biologically active rocks from a parent solar system that
impact the surfaces of (potentially habitable) terrestrial planets in
other solar systems is about 0.001 to 1.6 per cluster. With these
odds, biological transfer within a typical birth aggregate is quite
possible.  

The discussion thus far determines the likelihood of biologically
active rocks being transferred from a living solar system to the
surface of a potentially habitable planet in another system. In
practice, however, only a fraction $f_{seed}$ of these rocks will lead
to seeding of the new world. The difficulties associated with
atmospheric entry, crashing onto the surface, and the necessity of
landing in a nutrient rich location will lead to many failed attempts.
This additional probability factor, which cannot be calculated within
the scope of this paper, must be folded into any global assessment of
the odds of lithopanspermia.

These results are summarized in Figure 3, which shows the expected
number of lithopanspermia events $N_{\rm lps} = \tau \fimp N_B$ as a
function of mean ejection speed $\langle \veject \rangle$.  In this
context, the velocity dispersion that determines the capture cross
sections (see Figure 2) is given by the maximum of the mean ejection
speed $\langle \veject \rangle$ and the cluster velocity dispersion
$\vcluster$. The effects of varying the cluster size $\nstar$ are
illustrated by the three curves (for $\nstar$ = 100, 300, and 1000).
At low velocities $\veject < \vcluster$, the ejected rocks are bound
to the cluster and the expected number of lithopanspermia events is
maximized; in this limit, larger clusters produce higher numbers
$N_{\rm lps}$ of expected events.  At high speeds $\veject >
\vcluster$, the ejected rocks are unbound and generally pass through
the cluster only once.  In this limit, the effects of larger $\nstar$
(which increases the density of target solar systems) are nearly
canceled by the effects of increasing the cluster size $R$, so the
optical depth for scattering interactions (and hence $N_{\rm lps}$) is
independent of cluster membership $\nstar$. Although equations
(\ref{eq:taulow}) and (\ref{eq:tauhigh}) depict the transition between
the low and high velocity regimes as a step function, the actual
transition will not be as sharp. At intermediate ejection speeds
$\veject \sim 2-3$ km/s, some rocks will remain in the cluster for
several crossing times; in addition, some systems ($\sim10$ percent)
reside in the cluster core where the gravitational potential is deeper
than average, perhaps by a factor of $\sim10$ (Binney and Tremaine,
1987), so higher ejection speeds are required for the rocks to become
unbound. As a result, the transition from the low speed limit to the
high speed limit has been smoothed out in Figure 3 (with a transition
width of 1 km/s).

\subsection{Long Term Biological Transformation of a Cluster} 

The discussion thus far has focused on the first 10 Myr of evolution,
comparable to (but somewhat less than) the expected lifetimes for
embedded groups and clusters.  Although the majority of clusters will
disperse after 10 -- 20 Myr (e.g., Lada and Lada, 2003), some fraction
(roughly 10 percent) will remain bound for longer periods of time (100
-- 500 Myr). In such environments, life has more time to spread
throughout the cluster.  This subsection considers the general
transformation of a cluster from a nonliving state to one in which all
of the solar systems support life. Although most stellar aggregates
will not live long enough to complete this transition, this process is
important in long-lived clusters.

We assume here that the total number of stars $\nstar$ remains
constant over the time span of interest. Let $N_D$ be the number of
nonliving solar systems and $N_L$ be the number of living ones ($N_D +
N_L = \nstar$).  The infection rate $\Gamma$ -- the rate at which
living systems can transfer life bearing rocks to terrestrial planets 
in nonliving systems -- is given by 
\be 
\Gamma = \sigbar v n_B \, , 
\ee 
where $n_B$ is the number density of life bearing rocks in 
the cluster at a given time. The transition from nonliving 
to living systems is described by the differential equation 
\be 
{d N_D \over dt} = - \Gamma N_D = - \sigbar v n_B N_D \, . 
\label{eq:diff1} 
\ee 
The number density of life bearing rocks $n_B$ depends on the number
of living systems in the cluster and on how long they have been alive. 
If every living system provides $\gamma \approx$ 15 rocks/yr to the 
cluster, the population of life bearing rocks obeys the equation  
\be 
{d n_B \over dt} = \gamma n_\star (1 - f_D) \, , 
\label{eq:diff2} 
\ee 
where $f_D \equiv N_D/\nstar$ is the fraction of nonliving 
solar systems. We can combine equations (\ref{eq:diff1}) 
and (\ref{eq:diff2}) to obtain 
\be 
{d \over dt} \Bigl( {1 \over f_D} {d f_D \over dt} \Bigr) = 
- n_\star \sigbar v \gamma (1 - f_D) \, . 
\ee 
The time scale $t_0$ on which the fraction $f_D$ (and hence 
$f_L = 1 - f_D$) evolves is given by  
\be 
t_0 \equiv (n_\star \sigbar v \gamma)^{-1/2} \, \approx  
6.3 {\rm Myr} \, (\nstar/300)^{1/4} \sigwig^{-1/2} \, 
(v/1 {\rm km s}^{-1})^{-1/2} \, , 
\ee 
where we have used $\fimp = 10^{-4}$ and scaled the result using 
typical values in the second (approximate) equality. The fraction of 
nonliving systems in the cluster thus obeys the equation 
\be 
{d \over d \twig} \Bigl( {1 \over f_D} {d f_D \over d \twig} 
\Bigr) = - (1 - f_D) \, , 
\ee 
where the dimensionless time $\twig = t/t_0$. 
The solution for the fraction $f_L = (1 - f_D)$ of living systems as 
a function of dimensionless time is shown in Figure 4. As formulated
here, the solution can only depend on the initial condition $f_L(0)$,
where we take the starting time to be when the first solar system in
the group develops life. Figure 4 shows the solutions for three
starting conditions, $f_L(t=0)=1/\nstar$, with $\nstar$ = 100, 300,
and 1000. In smaller groups, the fraction $f_L$ of living systems is
larger at the initial time, so the curve for $f_L(t)$ increases faster
for smaller $\nstar$. Figure 5 shows the evolution in terms of
physical variables -- the larger clusters produce many more living
systems in the end (as expected), although the times scales are
somewhat longer. Notice that for typical values of $\nstar$, the time
required for a cluster to make the transition to fully living $(f_L
\to 1)$ is longer than the expected lifetimes of these systems. 

\subsection{Time scales for lithopanspermia in clusters} 

The basic scenario for lithopanspermia in clusters is outlined above.
To complete the discussion, we must examine the time scales involved
in the process. These time scales should be compared to the cluster
lifetimes, which are expected to be of order 10 -- 100 Myr, with
considerable variation (Lada and Lada, 2003; Porras et al., 2003).

The time required for biologically active rocks to be ejected from the
parent solar system is probably the most important bottleneck in the
process. For example, previous work has shown that the time required
for Martian ejecta to be removed from a solar system like our own is
typically 30 -- 50 Myr (Melosh, 2003; Gladman et al., 1996), where the
minimum time for ejection is about 4 Myr. The median ejection time
falls to only a few Myr when Jupiter is placed at the orbit of Mars;
extrapolation of Figure 3 from David et al. (2003) indicates that such
a solar system architecture would allow the Earth-like planet to
remain stable for (roughly) 200 Myr, longer than typical cluster
lifetimes ($\sim$100 Myr) and long enough for biological transfer to
occur. Since solar systems are expected to have a wide variety of
architectures, the time required for biologically active rocks to be
ejected will vary greatly from system to system. In some cases, the
rocks may not be ejected before the biologically active system leaves
the cluster, and the lithopanspermia process will suffer from an added
inefficiency.

The time required for ejected rocks to travel from system to system is
relatively short (compared to the other times scales of interest). The
ejection speeds lie in the range 1 -- 10 km/s and the cluster size is
of order 1 pc. The corresponding travel times thus fall in the range
0.1 -- 1 Myr, short enough not to be an issue.

Erosion poses yet another potential problem facing biologically active
rocks in space. For example, in the present Zodiacal cloud, a
meter-sized rock can be eroded on the relatively short time scale of
0.02 -- 0.23 Myr, where this time scale depends sensitively on the
relative speed between the rocks and the eroding medium (see Napier
2004). This time scale estimate is shorter than the time required for
rocks to be ejected from the solar system. Since younger systems
contain even more gas and can drive more severe erosion, this effect
introduces another inefficiency into the panspermia process.

Lithopanspermia involves another rather long time scale. After a rock
is captured by another solar system, a great deal of time passes
before the rock lands on the surface of a terrestrial planet in the
new system. The distribution of these time scales has been calculated
previously (Melosh, 2003). Most rocks are re-ejected (leading to the
small fraction $\fimp \sim 10^{-4}$) with a median time of about 60
Myr. Although this time scale is comparable to (or perhaps longer than) 
the cluster lifetime, it is not a major obstacle for panspermia: The
new solar system can seed itself over a longer time scale, as long as
the biologically active rocks are not stripped out of the system as
the cluster disperses.

Of course, the most important, and most uncertain, time scale is that 
required for life to develop in one of the member solar systems. 
Unfortunately, this time scale remains largely unknown.  

\section{CONCLUSION} 

In this paper, we have reconsidered the hypothesis of lithopanspermia
in the context of young, clustered star forming regions. In this
environment, the conditions are different from those considered
previously (namely in the field) in several respects: [1] The density
of solar systems is much greater and the relative velocities are
lower; these qualities tend to enhance the transfer of any
biologically active rocks that are present. [2] The systems live in
the cluster for only 10 -- 100 Myr, so the time scales available for
transfer are shorter. [3] We also consider the fact that most stars
reside in binaries; binarity increases the cross sections for the
capture of passing meteroids, but also decreases the range of
parameter space available for habitable planets.  In the early
formative phases of the solar system, rocky debris and bombardment are
much more common, which also enhances transfer. However, the
background UV radiation fields are stronger in regions of clustered
star formation and hence biological cargo is in greater danger.

\subsection{Summary of results}  

[1] We have calculated the cross sections $\sigcap$ for rocky bodies
to be captured by binary star systems (which make up the majority of
solar systems). To determine each cross section, we use a Monte Carlo
technique to sample the range of binary parameters, interaction
geometries, and interaction speeds (for a given velocity dispersion
$\sigma_v$).  The resulting cross section is a steeply decreasing
function of the velocity dispersion (see Figure 2) and can be fit with
a simple analytic form (see equation [\ref{eq:fit}]). The resulting
capture cross sections can be used in a wide variety of other
applications.

[2] Within typical star forming regions, the probability of any given
ejected rock being recaptured by another solar system is relatively
low. The effective optical depth for this interaction is sensitive to
the ejection speed of the rock from its original solar system. Over
the expected range of clusters with $\nstar = 30 - 1000$, the capture
optical depth $\tau \approx 0.0003 - 0.0043$ in the limit where the
mean ejection speed is less than or comparable to the velocity
dispersion of the stars in the cluster. For a higher mean ejection
velocity $\langle \veject \rangle \sim 5$ km/s, the optical depth is
much lower, $\tau \approx 10^{-6}$.  In both limits, however, most
rocks ejected from forming solar systems are not recaptured.

[3] The total number $N_R$ of ejected rocks per solar system is large
and the number of capture events $\ncap$ per cluster is given by
$\ncap \approx N_R \tau$.  Given the order of magnitude estimate $N_R
\sim 10^{16}$, every cluster will experience billions to trillions of
capture events. Essentially all solar systems in a given birth
aggregate are enriched (polluted) by rocks from other solar systems.
In other words, the sharing of rocky material among young solar
systems is inevitable.

[4] The number $N_B$ of biologically active rocks ejected from a
living solar system is estimated to be $N_B \sim 10^7$ over the time
spent in the birth aggregate (with considerable uncertainty).  For
clusters in which at least one system develops life, the number
$\nbio$ of capture events for biologically active rock (per cluster)
is given by $\nbio \approx N_B \tau$. In the low speed limit, the
number of biologically active rocks captured (per cluster) is about
$\nbio \sim 16,000$. This value is a steeply decreasing function of
the rock ejection speed and reduces to $\nbio \approx 10$ (per
cluster) for the benchmark value $\langle \veject \rangle = 5$ km/s.
If a solar system gives rise to life within a birth aggregate, then it
is likely to transfer life bearing rocks to the other solar systems in
the aggregate. (Note that only a fraction of clusters will develop
life and thus be capable of spreading it through the mechanism
considered here).

[5] Only a fraction of the captured rocks that are biologically viable
will strike the surface of a terrestrial planet and thereby complete
the lithopanspermia process. Previous studies estimated this fraction
to be $\fimp \sim 10^{-4}$. In general, the number of lithopanspermia
events is given by $N_{\rm lps} \approx \fimp N_B \tau$. In the limit
of low ejection speeds, only one or two lithopanspermia events are
expected in a typical birth cluster.  For higher speeds, $\langle
\veject \rangle \approx 5$ km/s, the expected number of successful
lithopanspermia events per cluster is $\sim10^{-3}$ and the odds of
successful lithopanspermia are about 1 out of 1000. These general
results are shown in Figure 3. A related quantity is the effective
efficiency of lithopanspermia, which is relatively low, i.e.,
$\epsilon_{\rm lps} \equiv N_{\rm lps}/\nstar \approx 3 \times 10^{-6}
- 5 \times 10^{-3}$.

The numbers quoted here apply only to those clusters that develop life
and implicitly assume that all rocks that land on habitable planets
will spread life. Neither of these assumptions is guaranteed to
hold. In order to assess the global probability of lithopanspermia,
one must take into account that only a fraction $f_{cl}$ of clusters
will develop life while they remain intact, and only a fraction
$f_{seed}$ of rocks that land on suitable planets will be successful
in establishing life.

\subsection{External versus internal seeding} 

In any panspermia scenario, a key bottleneck is the origin of life in
the first place. Of course, if the spontaneous origin of life were
sufficently common, there would no need for any panspermia mechanism
to explain the presence of life, although biological transfer would
still be of interest. In the present context, an important issue is
whether life is more likely to arise spontaneously within a birth
cluster or be captured from the outside. In order to make a
quantitative assessment, we assume that life will arise with
probability $p$ for any given solar system within a time span of 10
Myr. Solar systems forming within a group or cluster (which remains 
intact for about 10 Myr) will thus create life with probability $p$. 
The probability $P_C$ of a birth aggregate with $\nstar$ systems
giving rise to life spontaneously is thus $P_C \approx \nstar p$.

For comparison, we must estimate the probability of the birth cluster
capturing life bearing rocks from the outside. The optical depth for
capture is given by $\tau = n \sigma_T \ell$, where $n$ is the number
density of life bearing rocks, $\sigma_T$ is the capture cross section
of the entire cluster, and $\ell= vt$ is the effective path length.
Since $v \sim 40$ km/s and $t \sim 10$ Myr, the path length $\ell \sim
400$ pc. In the solar neighborhood, the number density of stars
$n_\star \approx 0.04$ pc$^{-3}$. If we assume that solar systems are
biologically viable for $\sim$10 Gyr, then individual systems will
create life with probability 1000$p$. The density of solar systems
that give rise to life is $1000 n_\star p \approx 40 p$ pc$^{-3}$. If
each living solar system ejects 15 life bearing rocks per year (\S
2.3), the density of life bearing rocks is $n \sim 6 \times 10^{12} p$
pc$^{-3}$. The capture cross section for the entire cluster is
approximately $\nstar \langle \sigma (40 {\rm km/s}) \rangle$, where
the cross section for an individual solar system to capture a high
speed rock is $\langle \sigma (40 {\rm km/s}) \rangle \approx$ 3
AU$^2$ (see equation [\ref{eq:fit}]). The probability of the cluster
capturing life bearing rocks is thus $\tau = n \sigma vt \approx
180,000 \nstar p$ (per cluster). In order to compare with our estimate
for the spontaneous rise of life, we must take into account the fact
that only a fraction $\fimp$ of the life bearing rocks captured in
this manner eventually strike the surface of a terrestrial planet and
make a successful transfer. For the benchmark value $\fimp \sim
10^{-4}$, the probability of a cluster being seeded from the outside
is $P_C \approx 18 \nstar p$. These results suggest that a young
cluster is more likely to capture life from outside than to give rise
to life spontaneously. Once seeded, the cluster provides an
effective amplification mechanism to infect other members.

This formalism also provides an estimate of the probability that Earth
has transferred life to other solar systems. This issue is especially
pertinent because Earth is the one planet where we know that life did
develop. We consider the case of continued ejection of life bearing
rocks over the age of the solar system, so we use conditions relevant
to the field.  The optical depth for a rock (from Earth) being
captured by another solar system is given by $\tau$ = $n_\star \sigma
v (\Delta t)$, where $n_\star \approx 0.04$ pc$^{-3}$, $v \approx$ 40
km/s, $\Delta t \approx$ 4 Gyr, and $\sigma$(40 km/s) $\approx$ 3
AU$^2$.  With these values, the optical depth for capture is $\tau
\approx 4.6 \times 10^{-7}$. Following the same approximation scheme
developed earlier, we assume that the Earth ejects about 10 life
bearing rocks per year and thus ejects $N_B \approx 4 \times 10^{10}$
such rocks during the time over which life has existed. Putting these
two results together, we find that $N_B \tau / 2 \approx 9,000$ life
bearing rocks will be captured by other solar systems (where we have
included the factor of two to account for the fact that the rocks that
are ejected first have a longer travel time). Using the standard
fraction $\fimp \approx 10^{-4}$ for the number of captured rocks that
make their way onto the surface of a habitable planet, the expected
number of transfer events is about 0.9.  In other words, these results
suggest that life on Earth can be transferred to {\it one} other
habitable world in another solar system. This value reflects the
steady state transfer rate of life from our solar system to others. In
addition, our planet is thought to have experience a period of ``late
heavy bombardment'' from about 4.4 to 3.8 Gyr ago, when a large number
of additional life bearing rocks could be ejected into space. This
epoch implies an enhancement in the transfer rate (and will be the
subject of a forthcoming paper -- G. Laughlin, private communication).

\subsection{Discussion and future work} 

This paper shows that young star clusters provide an efficient means
of transferring rocky material from solar system to solar system. If
any solar system in the birth aggregate supports life, then many other
solar systems in the cluster can capture life bearing rocks.  Only a
fraction of these systems will feed biologically active rocks onto the
surfaces of terrestrial planets, however, so the odds of successful
lithopanspermia are low: In the limit of low speed ejecta, only a few
systems per cluster are expected to be biologically seeded through
this mechanism, although the efficiency is reasonably high (about
$\epsilon_{\rm lps} \approx$ 0.005). If the origin of life is
relatively common and if life bearing rocks can be ejected at low
speeds, then dynamical interactions in stellar birth clusters would
provide an effective mechanism for spreading life.

This paper has explored the possibility that young clusters can lead
to greater efficiency of panspermia. However, these cluster
environments also present additional hazards for the transfer of
biological material. One obstacle is the increased levels of radiation
at ultraviolet wavelengths. Previous work on this subject has shown
that the radiation fields in small clusters (those with a few hundred
members) are generally not strong enough to affect circumstellar disks
and planet formation (e.g., Adams and Laughlin, 2001; Adams et al.,
2004), although a definitive assessment of the consequences for
biological material remains to be done. Further, the radiation fields
increase steeply with increasing cluster size $\nstar$ so that
sufficiently large clusters will present substantial hazards.  Another
potential issue is that of supernovae. Since only about 3 out of every
1000 stars are massive enough to end their lives in a supernova
explosion, such events are rare in small clusters (Adams and Laughlin,
2001). Furthermore, the most massive stars tend to live near the
cluster centers, so that the majority of stars will be $\sim1$ pc away
(see Adams and Myers, 2001).

In the discussion thus far, we have estimated the likelihood of
lithopanspermia events using conservative values for the input
parameters. The resulting odds of life being carried from solar system
to solar system are high enough to be tantalizing, but not high enough
to guarantee transfer. It is interesting to see what might happen with
more optimistic estimates. Consider the case of low ejection speeds so
that the velocity dispersion of solar systems in the cluster
determines the cross section, i.e., $\sigma_v \approx \vcluster
\approx 1$ km/s and hence $\sigwig \approx 1$. In this case, the
interaction optical depth $\tau = 3 \times 10^{-5} \nstar / \log
\nstar$.  Next we assume that the lower mass limit for spores to
survive is $m_B$ = 142 g (the mass of a baseball) so that $N_B \approx
1.7 \times 10^8$, and the efficiency of transfer is enhanced because
of the extreme collisional activity of planet formation so that $\fimp
\approx 10^{-3}$. We also assume that life bearing rocks that land on
suitable planets will be successful in spreading life so that
$f_{seed}$ = 1.  With these values, the expected number of
lithopanspermia events (per cluster) is $N_{\rm lps} \approx 5.1
\nstar / \log \nstar$ over the fiducial 10 Myr time scale. In this
limit, life can be transferred to every solar system in a group with
$\nstar = 100$. In a larger cluster with $\nstar = 1000$, 75\% of the
systems would become infected with life over the nominal 10 Myr time
period; however, larger clusters remain intact much longer and have
additional time to transfer biological material from system to
system. As a result, optimistic circumstances allow a cluster, once
biologically seeded, to transfer life to the majority of its solar
systems through the process of lithopanspermia.

To further our understanding of the lithopanspermia mechanism,
additional calculations must be performed. One important quantity is
the fraction $\fimp$ of captured material that falls onto the surfaces
of habitable planets. This paper follows previous authors and uses the
estimate $\fimp \approx 10^{-4}$. Starting with the locations of
captured rocks from the binary capture simulations, a large ensemble
of dynamical calculations should be performed to determine $\fimp$. A
related issue is that for sufficiently young solar systems,
biologically active rock can be captured while the planets are still
being assembled. During the planet formation epoch, rock-rock
collisions are common and spores could (in principle) be transferred
from rock to rock, leading to an enhancement in the effective value of
$\fimp$.  Another important quantity is the minimum mass necessary for
biological material to survive in space. In the setting of a young
cluster, the travel time is much lower than in the field (only $\sim1$
Myr compared to many Gyr), but the radiation fields are more
intense. And finally, the number of biologically active rocks ejected
by a given living planet should be estimated with greater precision.

\bigskip  
{}
\bigskip   

\centerline{\bf Acknowledgments}  
 
We would like to thank Greg Laughlin for useful conversations. We
thank both referees --- H. Melosh and B. Napier --- for many useful
comments that improved the paper.  FCA is supported at the University
of Michigan by the Michigan Center for Theoretical Physics and by NASA
through the Terrestrial Planet Finder Mission (NNG04G190G) and the
Astrophysics Theory Program (NNG04GK56G0).  DNS is supported by the
NASA Astrophysics Theory program.

\medskip 
{} 

\newpage 
\begin{figure}
\figurenum{1}
{\centerline{\epsscale{0.90} \plotone{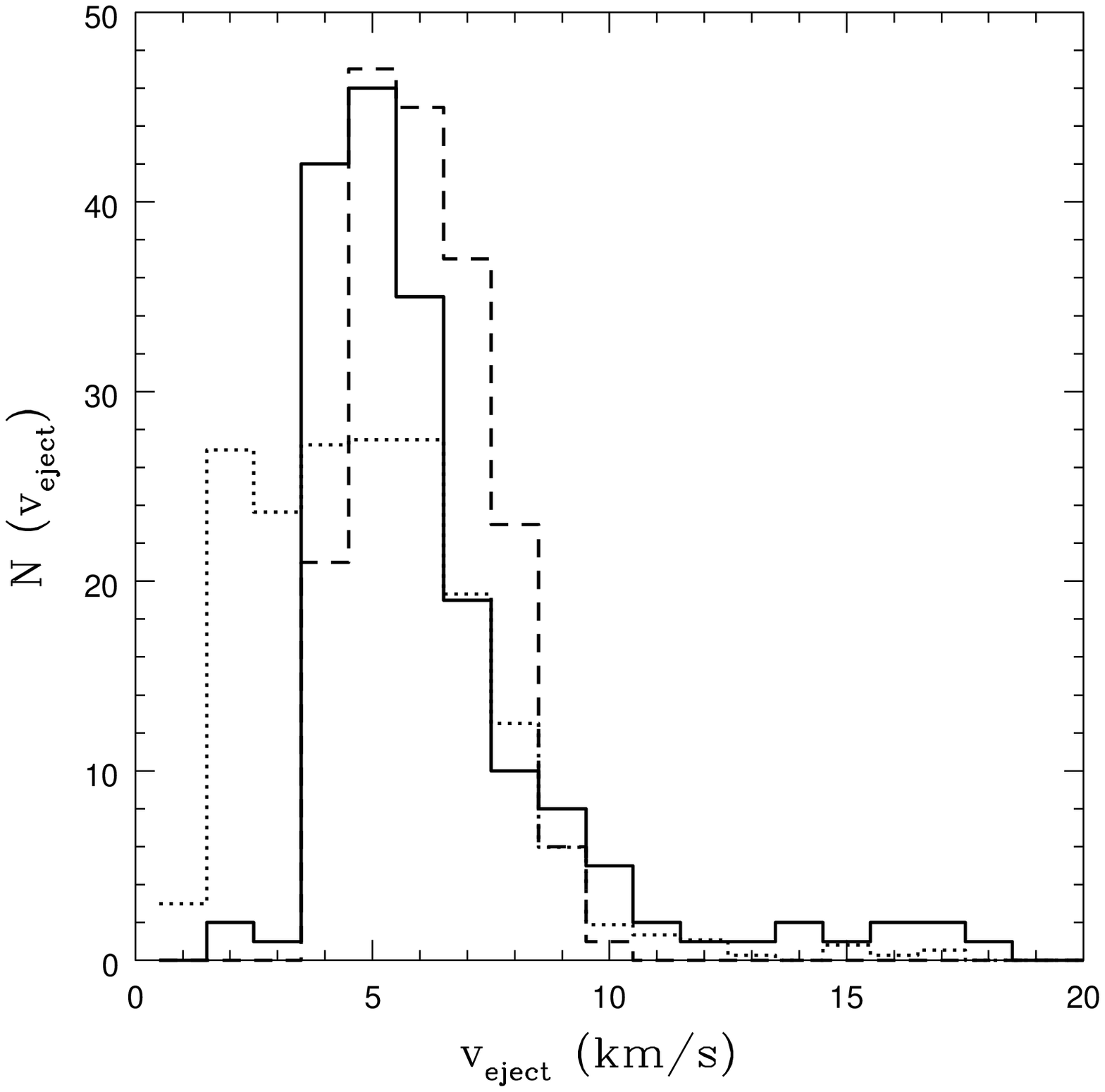} }}
\figcaption{Ejection speeds for rocky bodies removed from a solar
system. The solid curve shows the distribution of ejection speeds for
rocky bodies on Jupiter-crossing orbits. The dashed curve shows the
corresponding distribution of ejection speeds for rocky bodies on
Neptune-crossing orbits. The dotted curve shows the distribution of
ejection speeds for a stellar companion with mass $M_\ast$ = 0.1
$M_\odot$ and with semi-major axis $a$ = 42 AU (near the peak of the
binary period distribution).  All three distributions of ejection
speed shown here have the same normalization. }  
\end{figure}

\newpage 
\begin{figure}
\figurenum{2}
{\centerline{\epsscale{0.90} \plotone{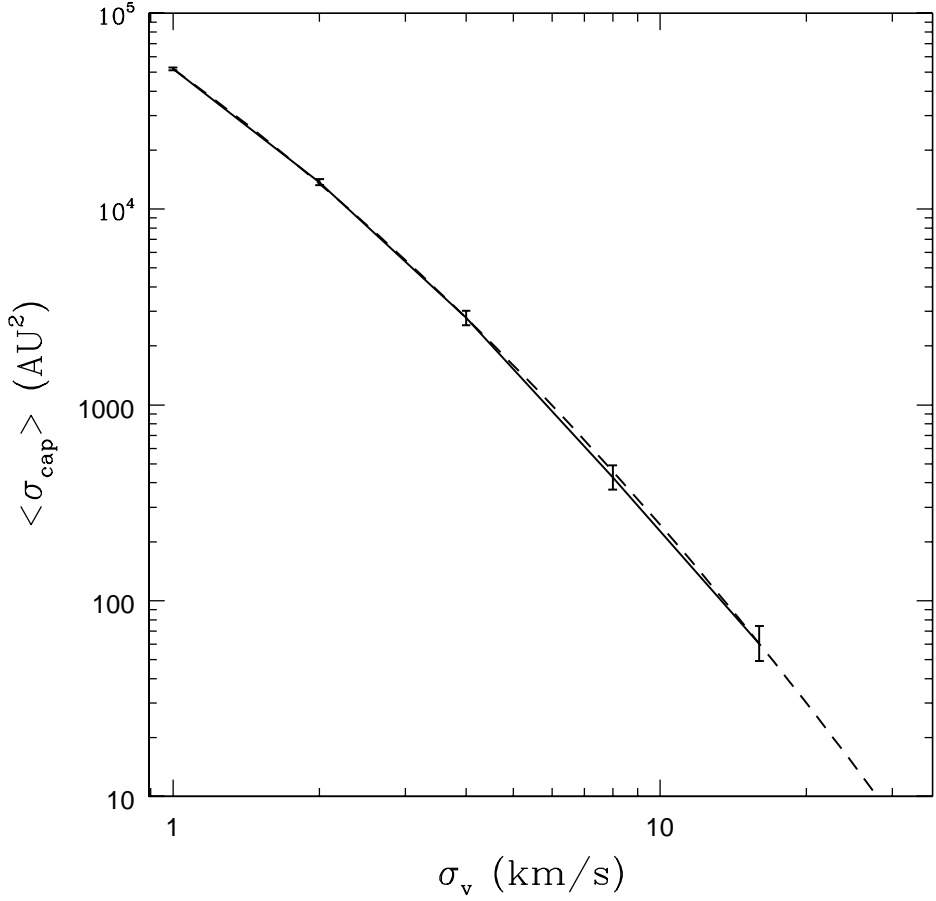} }}
\figcaption{Capture cross section for rocks interacting with binary
star systems. The cross section $\sigcap$ is shown as a function of
the velocity dispersion $\sigma_v$. Typical star forming groups and
clusters have $\sigma_v \approx 1$ km/s, corresponding to the left end
of the curve; stars in the field have $\sigma_v \approx 20 - 40$ km/s,
corresponding to the right side of the curve. Rocks that are scattered
out of solar systems display a range of ejection speeds (roughly
spanning the range shown here), depending on the location in their
solar system from which they are ejected.  The solid curve shows the 
result of our numerical simulations; the error bars depict the
one standard deviation errors resulting from the Monte carlo scheme
used to sample the input parameter space. The dashed curve shows an 
analytic fit to the cross section (see text). }   
\end{figure} 

\newpage 
\begin{figure}
\figurenum{3}
{\centerline{\epsscale{0.90} \plotone{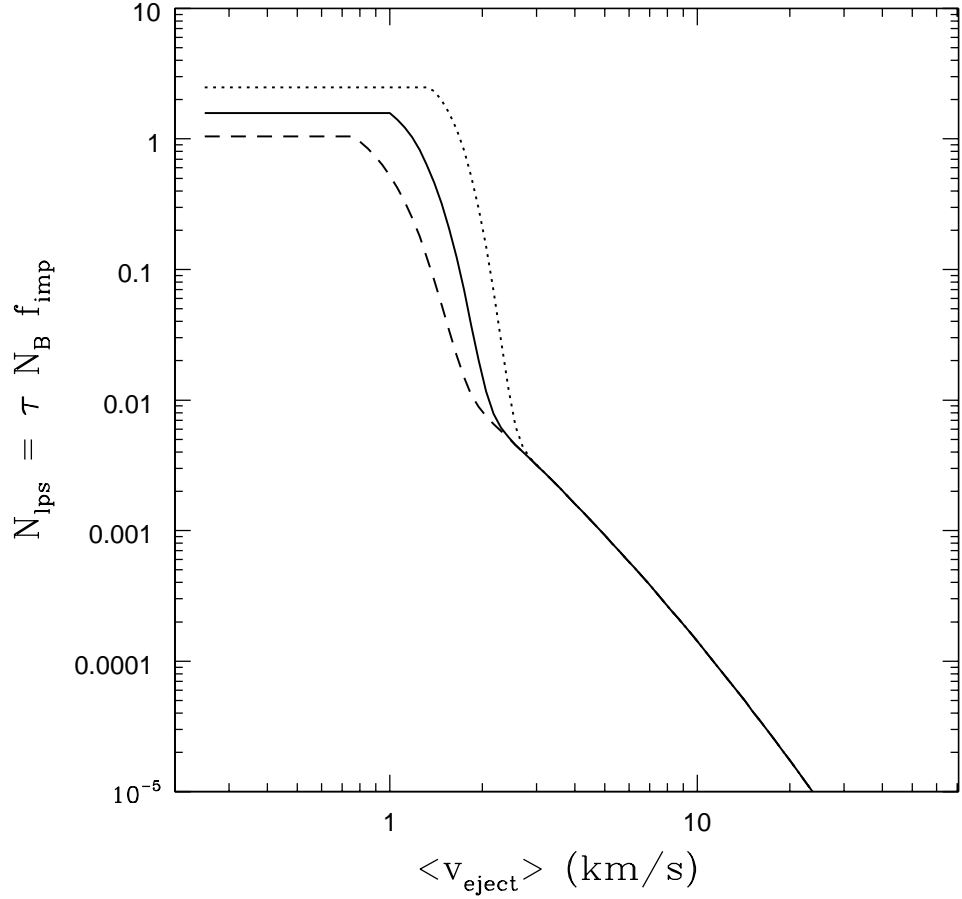} }}
\figcaption{Expected number of lithopanspermia events per cluster.
The three curves show the expected number of successful instances of
biological transfer as a function of the mean ejection speed $\langle
\veject \rangle$. The capture cross sections are a function of
velocity dispersion $\sigma_v$, which is set by the the maximum of the
mean ejection speed $\langle \veject \rangle$ and the cluster velocity
dispersion $\vcluster$. The solid curve shows the number of events
expected for a typical group/cluster with $\nstar$ = 300 members. The
dashed curve corresponds to $\nstar$ = 100 and the dotted curve
corresponds to $\nstar$ = 1000. }  
\end{figure}

\newpage 
\begin{figure}
\figurenum{4}
{\centerline{\epsscale{0.90} \plotone{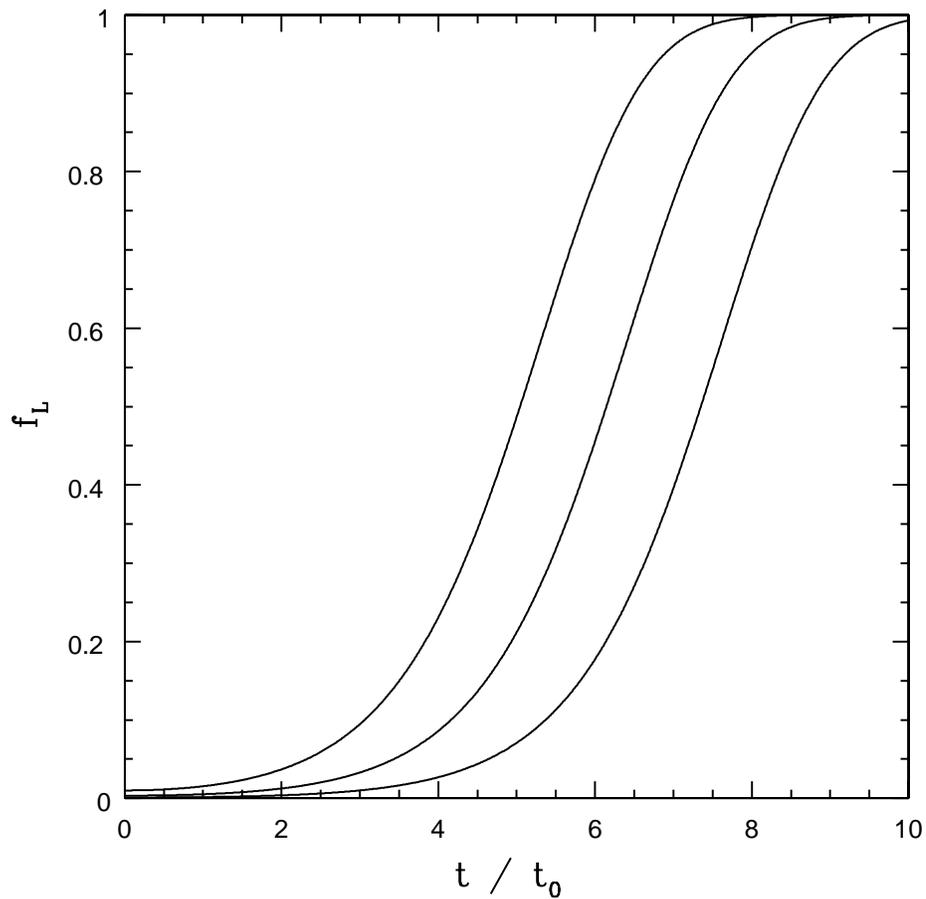} }}
\figcaption{Fraction of solar systems that contain life as a 
function of dimensionless time. The three curves show the result 
for different starting conditions, i.e., $f_L (t=0) = 1/\nstar$ 
with $\nstar$ = 100, 300, and 1000 (from left to right in the 
figure). }  
\end{figure}

\newpage 
\begin{figure}
\figurenum{5}
{\centerline{\epsscale{0.90} \plotone{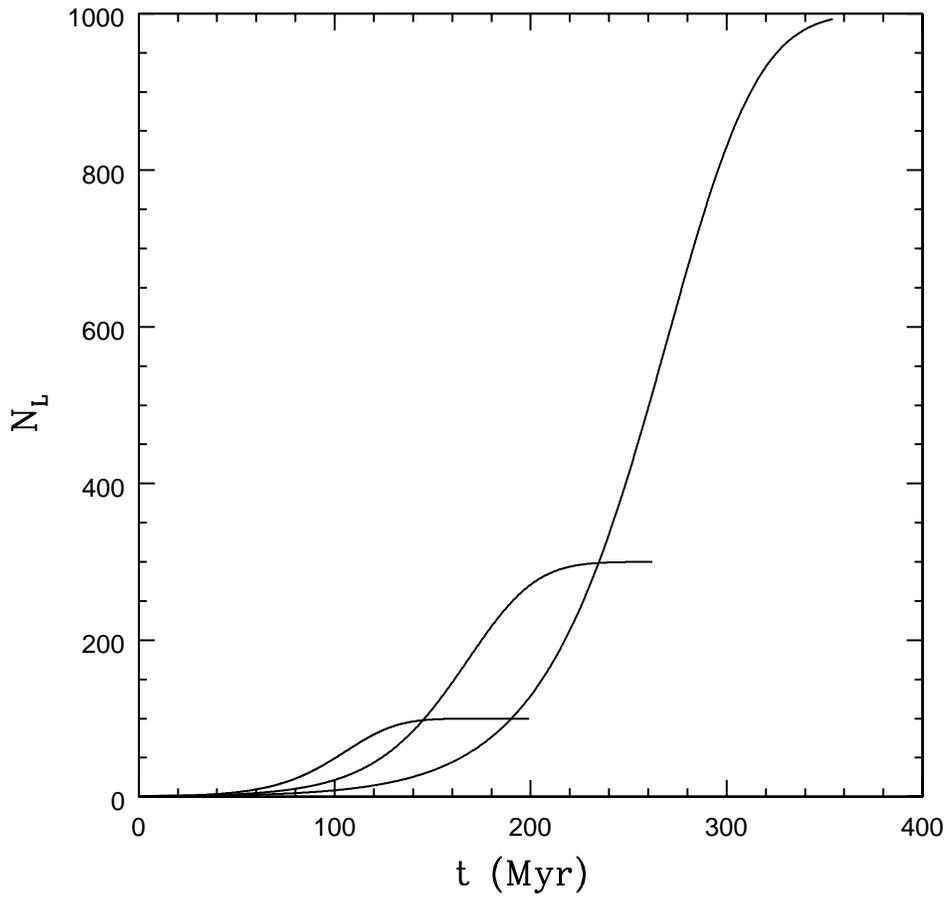} }}
\figcaption{Number of solar systems that contain life as a 
function of physical time (in Myr). The three curves show the result 
for different starting conditions, for clusters with $\nstar$ = 100, 
300, and 1000. Each curve asymptotically approaches $\nstar$ in 
the long time limit. }  
\end{figure}

\end{document}